\documentclass[twocolumn,preprint,amsfonts,amsmath,1p]{elsarticle}
\newcommand{\beq}[0]{\begin{equation}}
\newcommand{\eeq}[0]{\end{equation}}
\newcommand{\bw}[0]{\begin{widetext}}
\newcommand{\ew}[0]{\end{widetext}}
\newcommand{\bwn}[0]{\begin{widetext}\begin{eqnarray}}
\newcommand{\ewn}[0]{\end{eqnarray}\end{widetext}}
\newcommand{\beqn}[0]{\begin{eqnarray}}
\newcommand{\eeqn}[0]{\end{eqnarray}}

\newcommand{\no}[0]{\\\nonumber}
\newcommand{\np}[0]{{\it e.g. }}
\newcommand{\tzn}[0]{{\it i.e. }}
\newcommand{\proj}[1]{|#1\rangle \langle #1|}
\newcommand{\ket}[1]{|#1\rangle}
\newcommand{\bra}[1]{\langle #1 |}
\newcommand{\tens}[1]{\bigotimes _{#1=1} ^{k}}

\newcommand{\linia}[0]{\newline\noindent}

\newcommand{\proof}[0]{{\it Proof: }}

\newcommand{\kan}[0]{\Lambda}
\newcommand{\tr}[0]{\mathrm{tr}\:}

\newcommand{\flip}[0]{\mathbb{V}}
\newcommand{\jedynka}[0]{\mathbb{I}}
\newcommand{\swap}[1]{\mathbb{V}_{\ch_{#1}}}
\newcommand{\jed}[1]{\mathbb{I}_{\ch_{#1}}}

\newcommand{\akd}[1]{A_K ^{\ch_{1,2,\dots,#1}\dagger}}
\newcommand{\akdrugi}[1]{A_K ^{\ch_{#1+1,#1+2,\dots,2#1}}}
\newcommand{\akddrugi}[1]{A_K ^{\ch_{#1+1,#1+2,\dots,2#1}\dagger}}

\newcommand{\pplus}[1]{P_{+}^{\ch_{#1}}}
\newcommand{\esy}[1]{A_K ^{\ch_{1,2,\dots,#1}\dagger}\otimes A_K ^{\ch_{#1+1,#1+2,\dots,2#1}}}

\newcommand{\non}[0]{\nonumber\\}

\newcommand{\gj}[1]{\bigotimes _{#1 \in G}}

\newcommand{\ron}[0]{\varrho^{(n)}}
\newcommand{\gej}[0]{|{G}|}

\newtheorem{observation}{Observation}
\newcommand{\app}[1]{\renewcommand{\theequation}{#1-\arabic{equation}}
\setcounter{equation}{0}
\subsection*{Appendix #1}}
\def\id{\mathcal{I}}

\def\ch{{\cal H}}
\def\ce{{\cal E}}
\def\cd{{\cal D}}
\def\cn{{\cal N}}
\usepackage{amssymb}
\journal{Physics Letters A}
\begin{document}
\title{Aspects of multistation quantum information broadcasting}
\author{Maciej Demianowicz\corref{cor1}}\ead{maciej@mif.pg.gda.pl}
\author{Pawe\l$\;$Horodecki}\ead{pawel@mif.pg.gda.pl}
\cortext[cor1]{Corresponding author}
\address{Faculty of Applied Physics and Mathematics,
Gda\'nsk Technical University , 80-952 Gda\'nsk, Poland}
\begin{frontmatter}
\begin{keyword}
quantum information transmission \sep quantum broadcast channel \sep capacity theorem\sep capacity region \sep fidelity
\PACS 03.67.-a\sep 03.67.Hk
\end{keyword}
\begin{abstract}
We study quantum information transmission over multiparty quantum channel.
In particular, we show an equivalence of different capacity
notions and provide a multiletter characterization of a capacity region for a general quantum
channel with $k$ senders and $m$ receivers.
We point out
natural generalizations to the case of two-way classical communication capacity.
\end{abstract}
\end{frontmatter}
\section{Introduction}
Quantum channels have been in the field of interest since the early
stages of the development of quantum information theory. However,
the major progress in the domain have been achieved in the case of
quantum channels with single both
 sender and receiver, so--called
bipartite or single user channels
\cite{Schu9596,
BeDiSmoWoo96,BaNieSchu98,BaKniNie00,BeDiSmo97,BaSmoTer98,DevetakWinter}.
 Thorough investigations resulted in the quantum coding theorem which was
conjectured to exist in the form analogous to the form of
Shannon's theory \cite{Llo97,HoHoHo00,Sho02,De05}. Various aspects of single
user communication with assistance of different kind have been deeply analyzed
(see \cite{separations} for the hierarchic classification of capacities
in such scenarios). Nonadditivity of quantum channel capacity has been also
 reported \cite{SmiSmo09}. Recently some progress has been achieved
in the case of multiuser communication scenarios with both new
aspects and some generalizations of known results considered
\cite{DuHoCi04,YaDeHay05,DeHo06}.

In the paper we consider quantum information transmission over quantum channels. For a recent development
in classical or secret information capacities see \np \cite{Hast08,CzeHo08,LiWinGuoZuo09}.

The paper deals with a multiparty communication. First we
systemize  the notions of quantum channel capacity in this setup.
These are the generalizations of the ones from the single user
channel's theory \cite{KreWer03} and are known as entanglement
transmission, subspace transmission and entanglement generation.
We follow with the demonstration of the equivalence of this
scenarios. In the second part we provide a capacity theorem with a
simple proof for class of multiple antenna quantum broadcast
channels. Further we point out natural generalization to the case
of quantum capacity of a quantum channel assisted by two-way
classical side channel. Finally we summarize and discuss
our results.
\section{Background}
In this section we provide a short introduction into the area of quantum information transmission and a detailed background for further considerations.

A notion of a quantum channel introduced below is a standard mathematical notion used for the description of a physical disturbance to the quantum systems caused by the unavoidable interaction with an environment.
Our main concern will be a quantitative description of the issue of quantum information transmission through such channel.
Throughout the paper a shorthand notation $\psi\equiv \proj{\psi}$, logarithms are taken to base $2$.
%
\subsection{General view on the communication over quantum channels}
By the definition multiparty quantum channel with $k$ inputs and $m$ outputs ($km$-user channel, in short $km$-UC) is a completely positive trace preserving (CPTP) linear map $\Lambda$ acting from input density operators in $\mathcal{B}(\ch_{iCh})=\bigotimes_{j=1}^k \mathcal{B}(\ch_{iCh_j})$ to output density operators in $\mathcal{B}(\ch_{oCh})=\bigotimes_{i=1}^m
\mathcal{B}(\ch_{oCh_i})$, which in general can be of different dimensions. With this denotation
it can be formally written as $\Lambda: \mathcal{B}(\ch_{iCh})\rightarrow \mathcal{B}(\ch_{oCh})$.

We will consider a situation in which spatially separated $k$ parties, denoted ${\bf A}=\{ {\bf A} _i \}_{i\in \mathcal{K}=\{1,2,\dots,k\}}$ and called Alicias,
wish to communicate in a quantum manner spatially separated $m$ parties, denoted ${\bf B}=\{{\bf B}_i\}_{i\in \mathcal{M}=\{1,2,\dots,m\}}$ and called Bobbys.
Quantum information
embodied
in quantum systems sent by members of $\bf{A}$ is physically altered
what is described by $km$-UC. An implicit assumption of both classical and quantum information theory is that Alicias and Bobbys  have at their dispose $n$ ($n\to\infty$) instances of such channel which is usually written as $\kan ^{\otimes n}$ (this contains the assumption that the channel is memoryless). We assume that both groups act cooperatively
\tzn they agree to follow some jointly determined protocol which
 goal is, using $\kan ^{\otimes n}$, to establish a nontrivial reliable quantum communication channel between specified nodes of the network capable of faithful quantum information exchange. We will use single indices from the set $G$ to specify all two--nodes connections in the network. Obviously the number of elements in $G$ is $km$, however we will use $|G|$ for this number as this will allow for more clarity. The set is further divided into subsets on senders' and receivers' side \tzn $G=\{ G^{(j)}\}_j$ (note that the division differs for the parties on both ends of the channel).

Due to the different
goals $\mathbf{A}$ and
$\mathbf{B}$ want to achieve
we have different definitions of capacities \tzn different
approaches
to the problem of information transmission which we review below.

\subsection{Review of quantum communication notions}
\subsubsection{Entanglement transmission}
We start our review  with a concept of entanglement
transmission \cite{Schu9596,BaNieSchu98}.

We define the quantum sources $\mathfrak{S}_i=(\ron_{A_i},\ch _{A_i}^{(n)})_{n\in \mathbb{N}}$, $i\in I$, to be the pairs of sequences of Hilbert spaces and block density matrices on them \cite{BaKniNie00}. To the sources we assign entropy rates which are defined through $R_S(\mathfrak{S}_i)\equiv \limsup_{n\to\infty}S(\ron)/n:=\mathfrak{R}^e_i$; $S$ stands for von Neumann entropy, $I$ is some set of indices.
It is assumed that
every $\ron_{A_i}$
is the part of the larger system
$(RA)_i$ in some pure entangled state, \tzn
$\ron_{A_i}\equiv\mathrm{tr} _{R_i}\Psi_{(RA)_i}^{(n)}$ with the purifying system $\bf{R}$
assumed to be out of control of the parties.
Note that we can always look at the density matrix in this way.

The following sequences of operations constitute the protocol: (i) Alicias' CPTP collective encodings $\mathcal{E}^{(n)}=\otimes_{j\in \mathcal{K}}\mathcal{E}_j^{(n)}$,
$\mathcal{E}_j^{(n)}:\mathcal{B}(\mathcal{H}^{(n)}_{{\bf A}_j})\rightarrow
\mathcal{B}(\mathcal{H}^{\otimes n}_{{iCh}_j})$, $\ch _{{\bf A}_j}^{(n)}=\bigotimes_{i\in G^{(j)}}\ch _{A_i}^{(n)}$,
(ii) Bobbys' collectively CPTP decodings $\mathcal{D}^{(n)}=\otimes_{j\in \mathcal{M}}\mathcal{D}_j^{(n)}$,
$\mathcal{D}_j^{(n)}:\mathcal{B}(\mathcal{H}^{\otimes
n}_{{oCh}_j})\rightarrow
\mathcal{B}(\mathcal{H}^{(n)}_{{\bf B}_j})$, $\ch _{{\bf B}_j}^{(n)}=\bigotimes_{i\in G^{(j)}}\ch _{B_i}^{(n)}$.
The protocol together with the in-between usage of the channel $\Lambda ^{\otimes n}$  results in a sequence of channels $\mathcal{N}^{(n)}_{\mathbf{A}\to\mathbf{B}}$.

One says that the sources  $\mathfrak{S}_i$, $i \in G$, can be sent successfully (reliably) if there exists a protocol for which  entanglement fidelity defined as
\beqn\label{global1}
F_e\left(\gj{i}\ron_{A_i}
,\mathcal{N}^{(n)}_{\mathbf{A}\to\mathbf{B}}\right)\equiv
\mathrm{tr}\left[
\mathcal{I}^{\mathbf{R}}\otimes\mathcal{N}^{(n)}_{\mathbf{A}\to\mathbf{B}}
\left(\gj{i}\Psi_{(RA)_{i}}^{(n)}\right)
\gj{i}\Psi_{(RA)_{i}}^{(n)}\right]\nonumber\eeqn
tends to one in the limit of large $n$.
The $|G|$--tuple of rates $\{\mathfrak{R}^e_i\}_{i\in G}$ is said to be achievable if there exist sources with rates
$\mathfrak{R}^e_i$ that can be sent reliably. {\it Quantum channel capacity  is defined  to be a closure of the set of all $km$-tuples of achievable rates.}
The entanglement transmission capacity region will be denoted by
$\mathcal{Q}_e$.
To prevent unreasonable
situations in which rates are infinite
we concentrate only on sources satisfying quantum asymptotic equipartition property (QAEP; see \cite{KreWer03}).

 If we take input states
to be maximally entangled
we arrive at the notion of
maximal entanglement transmission.
The measure of reliability is called channel fidelity and is denoted by $F_c$.
A symbol $\mathcal{Q}_m$ will be used for the capacity region.

The fidelity used above is called global. As shown in
\cite{DeHo06} global fidelity is equivalent
to so called local ones  (\tzn convergence in global fidelity implies convergence in all local
fidelities and {\it vice versa}) which
are defined by ($i \in G$)

\beqn\label{local1}
F_e^{(i)}\left(\gj{i}\ron_{A_i}
,\mathcal{N}^{(n)}_{\mathbf{A}\to\mathbf{B}}\right)\equiv
\mathrm{tr}\left[ \left[\mathrm{tr}
_{\mathbf{RB}\setminus (RB)_i }
\mathcal{I}^{\mathbf{R}}\otimes\mathcal{N}^{(n)}_{\mathbf{A}\to\mathbf{B}}
\left(\gj{i}\Psi_{(RA)_{i}}^{(n)}\right)\right]
\Psi_{(RA)_{i}}^{(n)}\right],\nonumber\eeqn
where the partial trance means we trace out all the systems except
$(RB)_i$. We adopt the convention in which one of the arguments
of the fidelity is not the purification of $\varrho$ but only $\varrho$ itself. This is because
fidelities do not depend upon specific purification.
In the paper we also make use of
group fidelities, which are the ones with the specified
significant users traced out (obviously
they are equivalent to local and global fidelities).
These are denoted by $F^{[\mathcal{G}]}$
with $\mathcal{G}$ being any subset of G.
 We often omit one or both arguments of fidelities and  freely write
$F(\kan)$ or $F$ with proper superscripts causing no confusion as the arguments are
clear from the context. Absence of  superscripts  means we are
considering global fidelities.
This also concerns other fidelities considered further.

The definition of capacity region is general and is the same in all notions of capacity.
\subsubsection{Subspace transmission}
In the scenario of subspace transmission \cite{BeDiSmo97}
Alicias and Bobbys wish to transmit arbitrary pure states drawn from some Hilbert spaces.
One says that the sequence of  Hilbert spaces $\ch_{A_i}^{(n)}$, $i\in G$,  can be transmitted reliably if Alicias and Bobbys can use the protocol in such a manner that
minimum pure state fidelity defined as
\beqn
\label{purestate1}
F_s\left(
\gj{i}\ch_{A_i}^{(n)},\mathcal{N}^{(n)}_{{\mathbf{A}\to
\mathbf{B}}}\right)\equiv \min _{\gj{i}\ket{\psi_{A_i}^{(n)}}\in\gj{i}\ch_{A_i}^{(n)}}
\mathrm{tr}\left[ \mathcal{N}^{(n)}_{{\mathbf{A}\to
\mathbf{B}}}\left(\gj{i} {\psi^{(n)}_{A_i}}\right)
\gj{i}\psi ^{(n)}_{A_i}\right]
\eeqn
tends to one in the limit of large $n$. The $|G|$--tuple of rates $\{\mathfrak{R}^s_i\}_{i\in G}$ is said to be achievable if there exist sequences of Hilbert spaces $\ch_{A_i}^{(n)}$, $i\in G$, with $\limsup_{n\to\infty}(\log\dim
\ch_{A_i}^{(n)}/n)=\mathfrak{R}_i^{s}$ which can be sent reliably. Capacity region  is here denoted by $\mathcal{Q}_s$.

A similar scenario
arises when we choose average
fidelity as the reliability measure, {\it i.e.}
\beqn \bar{F}_{s} \left( \gj{i}\ch_{A_i}^{(n)},\mathcal{N}^{(n)}_{{\mathbf{A}\to
\mathbf{B}}} \right)=\int\Pi_{i\in G}\mathrm{d}\ket{\psi _{{A_i}}^{(n)}} \mathrm{tr}\left[ \mathcal{N}^{(n)}_{{\mathbf{A}\to
\mathbf{B}}}\left(\gj{i} {\psi^{(n)}_{A_i}}\right)
 \gj{i}\psi ^{(n)}_{A_i}\right]\nonumber,
\eeqn
where the integral is to be understood as $\int \mathrm{d}\ket{\psi} f(\ket{\psi})=\int \mathrm{d}U f(U\ket{\psi_0})$ with arbitrary $\ket{\psi_0}$ and RHS integral over all unitaries chosen according to the Haar measure on the subspace of interest.
Quantities which are averaged are called pure state fidelities (pure state fidelity for a state $\varphi$ and the channel will be denoted $F_s (\varphi,\kan)$).  Rates are defined as above and we use a denotation $\bar{\mathcal{Q}}_s$
for the capacity region.
\subsubsection{Entanglement generation}
Last considered here is the entanglement generation introduced in
\cite{De05}. The goal is to produce maximally entangled states between parties, $i\in G$. The first step of the protocol is replaced now by the preparation of a pure state
$\tens{j}\Psi_{({\bf AA'})_{j}}^{(n)}$, $\Psi_{({\bf AA'})_{l}}^{(n)}\in\ch_{\bf{A}_l}\otimes \ch_{iCh_l}^{\otimes n}$ (there is no further preprocessing ) as the input to the channel. The sequence arising from the protocol and the channel $\mathcal{N}^{(n)}_{\mathbf{A'}\to\mathbf{B}}$ is the concatenation of only the action of a channel and the decodings.
Generation of some fixed  $\ket{ \Phi_{d_{i}^{(n)}(AB)_{i}}^{(+)}}=
1/\sqrt{{d_{i}^{(n)}}}
\sum_{\gamma=0}^{d_{i}^{(n)}-1}\ket{\gamma_{A_i}}\ket{\gamma_{B_i}}$ with a given protocol
is said to be reliable if
entanglement generation fidelity defined as

\beqn F_g\left(\bigotimes_{i\in G}\Phi_{d_{i}^{(n)}(AB)_{i}}^{(+)}
,\mathcal{N}^{(n)}_{\mathbf{A'}\to\mathbf{B}}\right)\equiv
\mathrm{tr}\left[ \mathcal{I}^{\mathbf{A}}\otimes\mathcal{N}^{(n)}_{\mathbf{A'}\to\mathbf{B}}
\left(\tens{j}\Psi_{({\bf AA'})_{j}}^{(n)}\right)
\bigotimes_{i\in G}\Phi_{d_{i}^{(n)}(AB)_{i}}^{(+)}\right],\nonumber\eeqn
tends to one in the limit of large $n$.
One says that the $|G|$--tuple of rates $\{\mathfrak{R}^g_i\}_{i\in G}$ is achievable if there is a sequence of preparations allowing for reliable generation of maximally entangled states with $\limsup_{n\to\infty}(\log d_{i}^{(n)}/n)=\mathfrak{R}_i^g$. Capacity region is defined in analogy to the previous scenarios and is denoted by $\mathcal{Q}_g$.

There is no need to permit Alicias perform encodings as this would only mean that we let them prepare mixed instead of pure states at the beginning of the protocol, which does not provide us with substantially different communication scenario (cf. Section 3). However, when classical support comes into play (see the next subsection) it is reasonable to consider Alicias' operations
(preprocessing as well as operations during execution of the protocol).
To reduce the clutter we use the same denotation for both scenarios.

\subsection{Classical communication as a supportive resource}
So far we have not mentioned anything about additional resources
which may be used to enhance quantum transmission. Usually we let the
parties
share entanglement,
randomness, classical secret bits or communicate classically
(without any cost). In this paper we will be mainly concerned with a
special case of the last possibility, namely one-way forward
classical support denoted
with a superscript
$\rightarrow$, \np $\mathcal{Q}_s^{\rightarrow}$.
It is instructive to realize how the classical support fits into the quantum operation approach.
The connection is made by generalized measurements performed by Alicias. Learning upon the classical results $i$ ($i$ is a multiindex) of such measurements Alicias choose to perform $\ce _{i}$  (which are trace-decreasing, \tzn probabilistic, quantum operations) and inform Bobbys about the value of $i$ who can perform appropriate $\cd _{i}$ (which are trace preserving, \tzn deterministic operations).
It is now clear that entanglement generation in this scenario makes sense only if senders are allowed to operate on their parts, which was not the case in a zero-way regime. In a similar fashion we construct one-way backward  and two-way
protocols.
In case of
single user channels there is a well known result stating
uselessness of one--way forward classical support
\cite{BeDiSmoWoo96,BaKniNie00}. Recently the result has been generalized \cite{DeHo06}.
\subsection{Coherent information}
Here we recall one more quantity great importance of
which was conjectured long before its full recognition. It is the
coherent information \cite{BaNieSchu98,Llo97}, playing a role
similar to that of the mutual information in classical information
theory, defined as
$I_c(X>Y)_{\varrho^{AB}}$; $X=A,B$; $Y=B,A$.
We are not going into details concerning similarities and differences
between coherent and mutual information (for a recent
result see \cite{NatureMH}). We recall only one important feature,
namely quantum data processing inequality which states that coherent
information never increases in state postprocessing (operations $\mathcal{D}_{B\to B'}$ on $B$ side), {\it i.e.} $I_c(A>B)_{\varrho^{AB}}\ge
I_c(A>B')_{{\mathcal{D}_{B\to B'}(\varrho^{AB}})}$ \cite{BaNieSchu98}.
\section{Equivalence of capacity notions}
Now we turn to the first result of the paper. We show
that all introduced capacities are
the same in the sense that they give rise to the same capacity region.
One can notice that once
again the fundamental notion of teleportation finds its way to prove its usefulness.

\begin{observation} For 
multiparty quantum channel
it holds
$\mathcal{Q} _{g}=\mathcal{Q} _{m}=\mathcal{Q}
_{e}=\mathcal{Q} _{s}=\bar{\mathcal{Q}} _{s}
$.
\end{observation}
{\it Remark}: The problem of equivalence of different capacity
notions in case of a multiple access channel was considered in
\cite{YaDeHay05}. Here, as in \cite{DuHoCi04} and \cite{DeHo06},
we consider the most general scenario with $k$ senders and $m$
receivers. For bipartite case see \cite{KreWer03}.\newline\proof
({$\mathcal{Q}_e=\mathcal{Q}_s$)\label{rownowaznosc} This equivalence holds for sources satisfying
quantum asymptotic equipartition property.
For proofs see \cite{BaKniNie00} for bipartite and \cite{DeHo06}
multipartite case. For completeness of this paper we provide a revised multiparty proof in Appendix A.
\newline($\bar{\mathcal{Q}}_s=\mathcal{Q}_s$) Generalization of the technique
from Ref. \cite{KreWer03} provides us with the equivalence. From a
given reliable protocol
we construct a new classically supported protocol
which pure state fidelity equals average pure state fidelity of the
original one. Uselessness of classical side channel finishes the
proof. For details see the Appendix B.
\newline
($\bar{\mathcal{Q}}_s= \mathcal{Q}_m$) In the Appendix C we prove the
generalization of the formula from Ref. \cite{HoHoHo99} connecting
average fidelity with channels fidelity which with its local counterparts gives
the desired. In particular, for global average fidelity we have
 \beqn\label{averagechannel}
\bar{F_s}=\frac{1}{D_{+}}\left(D F_c+\displaystyle\sum _{j\in G}\frac{D}{d_j}F_c^{[G\setminus \{j\}]}\right.&+&\left.\sum _{i,j\in G;i\ne j}\frac{D}{d_i d_j}F_c^{[G\setminus \{i,j\}]}\right.+\no &&\left.\sum _{i,j,k\in G;i\ne j\ne k}\frac{D}{d_i d_j d_k}F_c^{[G\setminus \{i,j,k\}]}+\dots +1\right)
,\eeqn
where $d_i=\dim \ch _{A_i},D=\Pi_{i\in G}d_i,D_{+}=\Pi _{i\in G}(d_i+1)$.
The formula implies that in the limit of large dimensions average fidelity tends to the channel fidelity, \tzn
$\lim_{d_1,d_2,\dots,d_k\to\infty} \bar{F}_s=F_c.$
 If average fidelity is close to one then all channel smaller group fidelities are also high. So maximal entanglement transmission and average subspace transmission are equivalent. For the details of the derivation of (\ref{averagechannel}) see the Appendix C.
\newline ($\mathcal{Q}_{m}\subseteq\mathcal{Q}_g$) Consider a
protocol for sending maximal entanglement. Encoding of $i$th
sender results in some density matrix, which we can consider as a
mixture of pure states,
which, by convexity argument, means that for at least one
component of the mixture
we could achieve reliable transmission without the
necessity of encoding. Consequently it implies existence of a
protocol for generating entanglement with the rate at least as
good as for transmission of it. Naturally we also have $\mathcal{Q}_{m}\subseteq\mathcal{Q}_g ^{\rightarrow}$.
\newline($\mathcal{Q}_g\subseteq\mathcal{Q}_s$)
Generated entanglement can be used to perform teleportation
with high fidelity. In this way we have
$\mathcal{Q}_g\subseteq\mathcal{Q}_s^{\rightarrow}$ (by the same argument $\mathcal{Q}_g^{\rightarrow}\subseteq\mathcal{Q}_s^{\rightarrow}$ holds). The procedure
uses forward communication, which, as stated previously,
is useless
\tzn
$\mathcal{Q}_s^{\rightarrow}=\mathcal{Q}_s$. This inclusion is closely related to the problem of constructing a quantum error correction code from the distillation plus teleportation protocol
\cite{BeDiSmoWoo96}.
$\blacksquare$
\newline
The above results immediately imply that $\mathcal{Q}_g ^{\rightarrow}=\mathcal{Q}_g$. In a similar manner one shows that also the remaining scenarios do not gain any advantage acquiring free classical communication.
For an interesting backward classical communication scenario
see \cite{Leu08}.
\section{Capacity regions}
\subsection{Capacity theorem}
We turn now to the second result of the
paper. Namely, we give a multiletter characterization of the capacity region of the general
$km$--user channel.

 In case of $k=1$ and $m \ge 2$ we obtain a broadcast channel capacity region; for $k\ge 2$ and $m=1$ we get a multiple access channel, which capacity region
was recently provided in Ref. \cite{YaDeHay05} and was shown to be better than presented below for finite number $n$.
When $k\ge 2$ and $m\ge 2$ these two scenarios coexist.

The Observation 2, which we state below,  concerns zero-way capacity
region equivalent to the one-way one. We prove the result in the entanglement generation scenario which
according to Observation 1 is equivalent to other ones.
\newline\textbf{Observation 2 (Capacity region of a $km$--user channel)} {\it  Zero(one) -- way capacity
region $\mathcal{Q}(\Lambda)$ of a general $km$--user
channel $\Lambda:\mathcal{B}(\mathcal{H}_{\bf{A'}})\rightarrow\mathcal{B}(\mathcal{H}_{\bf{B}})$,
($\bf{A'}=A'_1 A'_2 ... A'_k$,$\bf{B}=B_1 B_2 \dots B_m$) is
given by the closure of $\;$
$\bigcup_{n=1}^{\infty}\frac{1}{n}\widetilde{\mathcal{Q}}(\Lambda^{\otimes
n}), $
%
where $\widetilde{\mathcal{Q}}(\Lambda)$ is the union  of $km$--tuple of nonnegative rates
$\{\mathfrak{R}_i\}_{i\in G}$  satisfying $\mathfrak{R}_i < I_c(A_i>B_i)_{\varrho_{(AB)_i}},$
over all
$\varrho_{{\bf AB}}=\left(
\mathcal{I}^{\bf{A}}\otimes\Lambda_{\bf{A'}\to\bf{B}}\right)\left(
\bigotimes_{\omega\in\mathcal{K}}\Psi_{({\bf AA'})_{\omega}}\right)$
which $\varrho_{(AB)_i}$ arise from by tracing out all the systems besides
$i$--th one.}
\newline \proof
{\it (achievability)} Alicias produce $\left[\varrho_{{\bf AB}}^{(n)}\right]^{\otimes \tilde{n}}\equiv\left[\left(
\mathcal{I}^{\bf{A}}\otimes\Lambda^{\otimes n}_{\bf{A'}\to\bf{B}}\right)\left(
\bigotimes_{\omega\in\mathcal{K}}\Psi^{(n)}_{({\bf AA'})_{\omega}}\right)\right]^{\otimes \tilde{n}}$
and perform with Bobbys one--way hashing
protocol of Devetak and Winter \cite{DevetakWinter} on
$\left(\varrho^{(n)}_{(AB)_{j}}\right)^{\otimes \tilde{n}}$ which achieves asymptotically
entanglement generation rates
$\frac{1}{n}I_c(A_j>B_j)_{\varrho^{(n)}_{(AB)_j}}$. Since
forward communication
is useless the rates are achievable in zero--way communication.

Before we proceed
we recall a useful
lemma (see \cite{De05})\linia
{\it Lemma;-
For  states $\varrho^{\mathcal{AB}}$ and $\sigma^{\mathcal{AB}}$, of the same $d$ dimensions, with fidelity
$F(\varrho ^{\mathcal{AB}},\sigma ^{\mathcal{AB}})\equiv (\tr |\sqrt{\varrho ^{\mathcal{AB}}}\sqrt{\sigma
^{\mathcal{AB}}}|)^2:=1-f$ we have
$|I_c (\mathcal{A}>\mathcal{B})_{\varrho ^{\mathcal{AB}}}-I_c
(\mathcal{A}>\mathcal{B})_{\sigma ^{\mathcal{AB}}}|\le
4\sqrt{f}\log d +2.$}
\newline{\it
(converse)} Consider entanglement generation protocol achieving
rates $\mathfrak{R}^{g}_i=\limsup_{n\to\infty}\mathcal{R}_i^{g(n)}$, where $\mathcal{R}_i^{g(n)}:=\log
d_j^{(n)}/n$
We have  $F^{(i)}_g\left(\bigotimes_{j\in G} \Psi_{(\bf{AA'})_{j}}^{(n)}
,\bigotimes_{l\in \mathcal{M}}\mathcal{D}_l^{(n)}\circ\Lambda^{\otimes n}\right)=1-\eta_n$
with $\eta_n\rightarrow 0$ for $n\rightarrow \infty$. Now taking in the Lemma $\varrho ^{\mathcal{AB}}$ as
$\mathrm{tr} _{\mathbf{AB}\setminus
(AB)_i}\left(\mathcal{D}^{(n)}\circ\Lambda^{\otimes
n}
\left(\bigotimes_{j\in\mathcal{K}}\Psi_{(\bf{AA'})_{j}}^{(n)}\right)\right)
\equiv\tilde{\mathcal{D}}_i^{(n)}(\varrho_{(AB)_i})$
and $\sigma ^{\mathcal{AB}}=
\Phi_{d_{i}^{(n)}(AB)_{i}}^{(+)}$ we have the following
justified by the data processing inequality and the
Lemma:
$I_c (A_i>B_i)_{\varrho_{(AB)_i^{(n)}}}
\ge
I_c(A_i>B_i)_{\tilde{\mathcal{D}}_i^{(n)}(\varrho^{(n)}_{(AB)_i})}
\ge n\mathfrak{R}^{g(n)}_i-2-8\sqrt{\eta}n\mathfrak{R}^{g(n)}_i
\ge n(\mathfrak{R}^{g(n)}_i-\delta_{\eta})$ 
with $\delta
_{\eta}\to 0$ when $n\to \infty$. This concludes the proof since the claimed set is closed.
$\blacksquare$
\newline One can easily verify that the region does not require convexification (cf. \cite{YaDeHay05}).

\subsection{Generalization to the two-way quantum capacity regions}
In a sense the above regions were derived by extended reasoning of
\cite{HoHoHo00} in that it utilizes (apart from data processing
inequality) two elements: hashing inequality for entanglement
distillation \cite{DevetakWinter} and the fact that
forward communication does not improve quantum capacity
(\cite{DeHo06}, \cite{BaKniNie00}). So
it is natural to ask about
possibility of extending the present results to the case of
two-way communication as it was in \cite{HoHoHo00}. The answer is
positive. All the reasoning  leading to theorems above uses either
zero-way (encoding, decoding) or one-way protocols
(teleportation). As it was in bipartite case one can follow any
protocol achieving some fixed coherent information rates by
one-way protocol involving entanglement distillation and
teleportation. The above leads to the following simple
conclusions: {\it Observation 1 is valid for all capacities
if we involve two-way encoding-decoding procedure. Also capacity
regions provided in Observation 2 are true if only in a place of the
state we put arbitrary state that can be produced
with help of a quantum channel $\Lambda^{\otimes n}$ assisted by
two-way LOCCs.  Finally, the multiple access channel's capacity
region provided in \cite{NatureMH,YaDeHay05} can be extended to two-way case in such a manner.}

\section{Summary and Discussion}
We have rigorously defined entanglement transmission, subspace transmission and entanglement generations
in case of multiparty quantum channels, systematized known facts about equivalences of capacity
notions, and shown truthfulness of the above in case of any
multiuser communication scenario. Using this fact with the aid of
recently proved uselessness of unlimited forward classical communication we provided
capacity regions for  $km$-user channels, which special cases are
the broadcast, multiple access, and $k$--user channels.
It seems that further
improvements of the region providing better approximations for finite $n$ not involving some
assumptions about the specific channel may be difficult.
It would be also desirable to find single-letter characterizations for classes of channels. However, at this point this remains an open question.
Finally we have
pointed out elementary generalization of the results  to the case
of two-way capacity regions.
In future it would be interesting to study the gap between the
case of zero-way (one-way) case and the case  of
two-way supported quantum channel in a general
$km$-user scenario.

After having had completed the main part of this work (quant-ph/0603112) we have
become aware of the result of the Ref. \cite{broadcast}
(quant-ph/0603098) where broadcast channels were considered.

\section{Acknowledgments} Discussions with Michal Horodecki are acknowledged. This
work was prepared under  EC IP project SCALA. M.D. is supported by Ministerstwo  Nauki i Szkolnictwa
Wyzszego grant no. N N202 191734.
\section{Appendices}
\app{A}
For completeness of the paper  we recall, with some details refined, the proof of equivalence between entanglement
and subspace transmission \cite{BaKniNie00,DeHo06}.
First we prove that entanglement transmission implies subspace transmission \cite{simplify}.

We assume that
$F_e(\bigotimes_{i\in G} \varrho_{A_i},\kan) \ge 1-\eta$,
where $\varrho_{A_i}$ are the normalized density matrices of the transmitted QAEP sources $\mathfrak{S}_i$ projected onto their typical subspaces with $\dim\mathrm{supp} (\varrho_{A_i})=K_i$ (this projection does not decrease substantially the fidelity; \cite{BaKniNie00}). Consider the following strategy.
We find a  vector $\ket{\varphi_{A_1}^{(1)}}\in \mathrm{supp} (\varrho_{A_1})$ that minimizes fidelity of the state $\varphi_{A_1}^{(1)}\otimes\left( \bigotimes_{i \in G\setminus \{1\}}\ron_{A_i}\right)$; we will refer to this fidelity as to $F_{e,s}$ as this is of mixed type.
We construct an operator $\rho_{A_1}^{(1)}=\varrho_{A_1}-q_1^1\varphi_{A_1}^{(1)}$
taking $q_1^1$ as large as we can still protecting positive semi-definiteness of the operator. We can proceed with the same strategy until we reach a zero operator. By construction in step $k+1$ we obtain an operator of dimension one less than in step $k$ (we have removed one dimension from the support). What is more we get that $\{q_m^1,\varphi_{A_1}^{(m)}\}$ constitutes a pure state ensemble for $\varrho_{A_1}$, \tzn $\varrho_{A_1}=\sum_{m=1}^{K_1}q_m^1\varphi_{A_1}^{(m)}$. Let us assume that using this strategy we removed $d_1$ dimensions from the $K_1$-dimensional support obtaining a subspace $\ch_{D_1}$, from which for all states we have $F_{e,s}\ge 1-\gamma_1$. If we further denote $\alpha_1=\sum_{m=1}^{d_1}q_m^1$ and by $\varrho_{A_i}^{d_i}$ normalized density matrix with $d_1$ dimensions removed we can rewrite global entanglement condition as
$F_e \left(\left( \sum_{m=1}^{d_1}q_m^1\varphi_{A_1}^{(m)}+ (1-\alpha_1)\varrho_{A_1}^{d_1}   \right)\otimes \bigotimes_{i\in G\setminus \{1\}}\varrho_{A_i} ,\kan\right) \ge 1-\eta$.
By convexity of entanglement fidelity in the input density operator, \tzn $F_e(\sum_i p_i \rho_i, \Lambda)\le \sum_i p_i F_e (\rho_i,\Lambda)$, we get
$1-\eta \le (1-\gamma_1) \alpha_1 +(1 -\alpha _1),$
which gives $\gamma_1\le \eta /\alpha_1$.
%
Repeating the above procedure for the rest $i\in G\setminus\{1\}$ we obtain
$F_s(\bigotimes_{i\in G}\ch_{D_i},\kan)\ge 1-\gamma_{|G|}$, where $\gamma_{|G|}\le \eta/ \Pi_{i\in G}\alpha_i$. We thus obtained a factor by which the entanglement fidelity is decreased when we consider transmission of $\ch_{D_i}$.
Next we get the bounds for the dimensions of these subspaces. For all $i$ and $m$ we obviously have
$q_m^i \le \lambda_{max} (\varrho_{A_i})$, which due to the fact that $\varrho_{A_i}$ are normalized density operators restricted to the typical subspaces can be further bounded from above by $2^{-n (\mathfrak{R}_S(\mathfrak{S}_i) -\epsilon_i)}/(1-\delta _i)$ with the denominator being the probability of the projection onto the typical subspace.
All these combined gives
$D_i \ge (1-\alpha_i) |T_{\epsilon}^{i(n)}|,
 |T_{\epsilon}^{i(n)}|=(1-\delta_i) 2^{n (\mathfrak{R}_S(\mathfrak{S}_i)-\epsilon )}.$
This implies, with the initial assumption of having had arbitrarily taken $\varrho_i$ into account,
that the same region for entanglement transmission can also be achieved for subspace transmission.

Now let us move to another direction of implication. Using the technique of Ref. \cite{BaKniNie00,KnillLa97} we will show that if the product Hilbert space is reliably sent through the channel then product density matrix supported on the subspace of it can also be sent with high fidelity.
Suppose that $F_s (\bigotimes_{i \in G} \ch_{A_i},\kan)\ge 1-\eta$, \tzn $F_s (\bigotimes_{i \in G} \varphi_{A_i},\kan)\ge 1-\eta$ for all $\varphi_{A_i}\in \ch_{A_i}$, $i \in G$.
Writing the first state as a superposition of basis states \tzn $\varphi_{A_1}=\sum_{k}\sqrt{\lambda_k^1}\mathrm{e}^{\mathrm{i}\phi_k^1}\ket{k_1}$ and putting it to the above condition followed by averaging over phases, which does not decrease fidelity,
one obtains
\beqn
\bar{F}_s=&&
\hspace{-0.5cm}
\sum_{kl} \lambda_k^1 \lambda_l^1 \bra{k_1}\left(\bigotimes_{i\in G\setminus \{1\}}\bra{\varphi_{A_i}}\right)\kan \left(\ket{k_1}\bra{l_1}\otimes \bigotimes_{i\in G\setminus \{1\}}\varphi_{A_i}\right)\ket{l_1}\left(\bigotimes_{i\in G\setminus \{1\}}\ket{\varphi_{A_i}}\right)+\non
&&\hspace{-14mm}\sum_{km,k\ne m} \lambda_k^1 \lambda_m^1 \bra{m_1}\left(\bigotimes_{i\in G\setminus \{1\}}\bra{\varphi_{A_i}}\right)\kan \left(\ket{k_1}\bra{k_1}\otimes \bigotimes_{i\in G\setminus \{1\}}\varphi_{A_i}\right)\ket{m_1}\left(\bigotimes_{i\in G\setminus \{1\}}\ket{\varphi_{A_i}}\right).\nonumber
\eeqn
Direct calculation shows that the first term is the fidelity of the state $\varrho_{A_1}\otimes\left( \bigotimes_{i \in G\setminus \{1\}}\varphi_{A_i}\right)$, $\varrho_{A_1}=\sum_{k}\lambda_k^1 \proj{k_1}$, sent through the channel.  By the same arguments as in Ref. \cite{BaKniNie00} one can show that $F_{e,s}\left(\varrho_{A_1}\otimes\left( \bigotimes_{i \in G\setminus \{1\}}\varphi_{A_i}\right),\kan\right)\ge 1-\frac{3}{2}\eta$. We follow with the same strategy of averaging
which results in a bound for entanglement fidelity as follows $F_e \left(\bigotimes_{i\in G}\varrho_{A_i},\kan\right)\ge 1-(\frac{3}{2})^{|G|}\eta$.
To argue that the same capacity region for subspace
transmission is also achievable for entanglement transmission we take uniform density matrices on the transmitted spaces.

\app{B}

We follow modified strategy of Ref. \cite{KreWer03} to prove the desired equivalence.
We supplement the protocol with a specially constructed classical forward channel so that the new
channel is as follows \beqn
\cn^{\rightarrow}(\cdot)=\sum_{\vec{n}}\frac{1}{\Pi
_{i\in G}N_{i}}
\bigotimes_{i\in G}U_{n_{i}}^{\dagger}
\mathcal{N} \left(\bigotimes_{i\in G}U_{n_{i}} \left(\cdot\right)
\bigotimes_{i\in G}U_{n_{i}}^{\dagger}\right)
\bigotimes_{i\in G}U_{n_{i}}.\eeqn
The vector
$\vec{n}=(n_{1},n_{2},...,n_{\gej})$ represents
a classical message sent to receivers, $n_i$ are
taken from $N_{i}$ elements sets. At this moment we refrain
from specifying the sets of
unitary $U$.
Consider now pure state fidelity in our scenario
which, assuming that $\mathcal{N} (\cdot)=\sum_{j} A_j (\cdot)A_j ^{\dagger}$ and $N=\Pi_{i=1}^{\gej}N_i$, yields
\beqn
\label{fidelity}
F_s\left(\bigotimes_{i\in G}\psi_{A_i},\mathcal{N}^{\rightarrow}\hspace{-0.12cm}\right)\hspace{-0.1cm}=
\hspace{-0.1cm}\frac{1}{N}\sum_{j}\sum_
{\vec{n}}
\hspace{-0.1cm}\left(\bigotimes_{i\in G}\bra{\psi_{A_i}}U_{n_i}^{\dagger}\right)\hspace{-0.1cm}A_j\hspace{-0.1cm}
\left(\bigotimes_{i\in G}U_{n_i}\psi_{A_i}U_{n_i}^{\dagger}\right)\hspace{-0.1cm}A_j^{\dagger}
\hspace{-0.1cm}\left(\bigotimes_{i\in G}U_{n_i}\ket{\psi_{A_i}}\right)%
\nonumber
\eeqn
Now we ask whether we can
replace the sums with
integrals and
how should be the sets of $U$ be chosen if the answer is positive. We discuss these questions in what follows.
\newline \indent Define an operation
\beqn \label{chanreduced} \mathcal{N} _{\vec{n}\setminus n_1}^{(1)}
(\cdot)=\sum_{j}A_{\vec{n}\setminus n_1}^{(1)j}(\cdot)
(A_{\vec{n}\setminus n_1}^{(1)j})^{\dagger},\eeqn
where Kraus
operators are defined by the partial inner product
\beqn \label{inner}
A_{\vec{n}\setminus n_1}^{(1)j}=\left(\bigotimes_{i\in G\setminus\{1\}}
\bra{\psi_{A_i}}U_{n_i}^{\dagger}\right)A_j \left(\bigotimes_{i\in G\setminus\{1\}}
U_{n_i}\ket{\psi_{A_i}}\right).\eeqn
Eq. (\ref{fidelity}) then takes the
form
\beqn \label{fid}
F_s=\sum _{\vec{n}\setminus
n_1}\frac{N_1}{N}
\bra{\psi_{A_1}}\left(\frac{1}{N_1}\sum_{n_1}U_{n_1}^{\dagger}\mathcal{N}
_{\vec{n}\setminus n_1}^{(1)}(U_{n_1}\psi_{A_1}U_{n_1}^{\dagger})U_{n_1}\right)\ket{\psi_{A_1}}.
\eeqn
From the theory of unitary 2--designs \cite{Dan0506Aud06} we know
that in cases when we deal with $U(2^N)$ we have an equivalence
$\frac{1}{K}\sum_{k}U_k ^{\dagger}\mathcal{N} (U_k \varrho U_k
^{\dagger})U_k=\int \mathrm{d}U U^{\dagger}\mathcal{N} (U\varrho
U^{\dagger})U$
with suitable chosen $\{U_k\}$, which were shown
to be the Clifford group $\mathcal{C} _N$. We
can directly use this fact since here we deal with spaces of the proper dimensions. This follows
from the possibility of bounding the dimensions of the
transmitted spaces in the following manner $2^{l_n}\le {d_n}\le
2^{l_n+1}$ with $l_n\to\infty$ when $n\to\infty$ and restricting ourselves to the spaces of dimension from
the LHS of the first inequality which leaves
rates and fidelities unchanged. This turns Eq. (\ref{fid}) into
\beqn F_s= \sum _{\vec{n}\setminus n_1}\frac{N_1}{N}
\bra{\psi_{A_1}}\left(\int \mathrm{d}U_{n_1} U_{n_1}^{\dagger}\mathcal{N}
_{\vec{n}\setminus n_1}^{(1)}(U_{n_1}\psi_{A_1}U_{n_1}^{\dagger})U_{n_1}\right)\ket{\psi_{A_1}}. \eeqn
Now taking
back the step (\ref{chanreduced}) and applying analogous procedure to
the remaining $|G|-1$ states we obtain
\beqn F_s=\int \Pi
_{i\in G}\mathrm{d}U_{n_i}\bigotimes_{i\in G}\bra{\psi_{A_i}}U_{n_i}^{\dagger}\mathcal{N}
\left(\bigotimes_{i\in G}U_{n_i}\psi _{A_i}
U_{n_i}^{\dagger}\right)\bigotimes_{i\in G}U_{n_i}\ket{\psi_{A_i}}, \eeqn
which
is just the average pure state fidelity $\overline{F}_s$. By the
uselessness of classical  channel we conclude the
equivalence of capacities.

\app{C}

Here we prove Eq. (\ref{averagechannel}). In what follows
sub- and superscripts denoted with $\ch$ will indicate
spaces on which operators act, channel $\kan$ has Kraus operators $\{A_K\}_K$.
Using the approach from Ref. \cite{KreWer03} we arrive at:
\beqn\label{ogolny}
\bar{F}_s (\kan)= \bar{D}\sum _K\tr \esy{\gej} \gj{i} \;(\jed{i,\gej+i}+\swap{i,\gej+i}),\nonumber
\eeqn
where $\bar{D}=\Pi _{i\in G} d_i^{-1}(d_i+1)^{-1}$
which can be rewritten as
 \beqn \label{ogolnyrozpis}
&&\hspace{-0.8cm}\bar{F}_s (\kan)=\bar{D} \sum _K\tr \esy{\gej}
\left[\jed{1,2,\dots,\gej,\gej+1,\dots,2\gej}+\right.
\non &&\hspace{-0cm}\sum _{\pi} \left.(\displaystyle \frac{1}{ 1! (\gej -1)!}
\swap{1,\gej+1}\otimes\jed{G^{2}\setminus \{1,\gej+1\}}+
\frac{1}{\displaystyle  2! (\gej -2)!}\times
\right.
\\ && \left.
\hspace{-0.8cm}\swap{1,\gej+1}\otimes\swap{2,\gej+2}\otimes\jed{G^{2}\setminus \{1,2,\gej+1,\gej+2\}}
+\dots)+\swap{1,2,\dots,\gej,\gej+1,\gej+2,\dots,2\gej}\right],\nonumber
\eeqn
where the permutation $\pi$ permutes Hilbert spaces $\ch _{i,\gej+i}=\mathbb{C}^{d_i}\otimes \mathbb{C}^{d_i}$ and  $G^{2}=\{1,2,\dots,|G|,|G|+1,\dots,2|G|\}$. In general $\ch_{klm\cdots}\equiv \ch_k \otimes \ch_l \otimes \ch_m \otimes \cdots$ and $\flip_{\ch_{\vec{a},\vec{b}}}\ket{\phi}\ket{\psi}=\ket{\psi}\ket{\phi}$, $\ket{\phi}\in \ch_{\vec{a}}$, $\ket{\psi}\in \ch_{\vec{b}}$, $|\vec{a}|=|\vec{b}|$.
Now we will associate all the terms in the above with the proper channel fidelities.

Let us start with a calculation of global channel fidelity. We have
\beqn
F_c (\kan)=&&
\sum _K \tr \left[ \jed{1,2,\dots,\gej}\otimes \akdrugi{\gej} \left(\bigotimes _{i\in G} \pplus{i,i+\gej}\right)\right.\times \non &&\hspace{2cm}\times\left. \jed{1,2,\dots,\gej}\otimes \akddrugi{\gej} \left(\bigotimes _{i\in G} \pplus{i,i+\gej}\right)\right],\nonumber
\eeqn
where $P_+ =\proj{\Phi^{(+)}}$ is a maximally entangled state projector acting on $\mathbb{C}^{d}$. Now, after having used the following properties: $\tr A_{12}^{\Gamma_1}B_{12}^{\Gamma_1}=\tr A_{12}B_{12}$, $\tr (\jedynka _1 \otimes A_2 \varrho _{12} \jedynka _1 \otimes B_2)^{\Gamma_1}=
\tr \jedynka _{1}\otimes A_{2} \varrho_{12} ^{\Gamma _1} \jedynka _{1}\otimes B_{2}$, and $dP_+ ^{\Gamma}=\flip$ in the order they are quoted here,
we obtain
\beqn
F_c (\kan)= &&\frac{1}{\Pi _{i\in G} d_i ^2}\sum_ K\tr\left[\jed{1,2,\dots,\gej}\otimes \akdrugi{\gej} \left(\bigotimes _{i\in G} \swap{i,i+\gej}\right)\right.\times\non && \hspace{2cm}\times\left. \jed{1,2,\dots,\gej}\otimes \akddrugi{\gej} \left(\bigotimes _{i\in G} \swap{i,i+\gej}\right)\right],\nonumber\eeqn
which can further be rewritten as
\beqn
F_c(\kan)=\frac{1}{\Pi _{i\in G} d_i ^2}\sum_ K \tr \akd{\gej} \otimes \akdrugi{\gej},\nonumber
\eeqn
which, up to a constant factor, is the first term in the considered sum.

Let us now move to group fidelities. We will describe our procedure in details for group fidelities of order $\gej-1$ as the method easily generalizes. These will be denoted $F^{[G\setminus\{k\}]}$, where $k$ is the enumeration of connection which is traced out.  We give a method of calculation of $F^{[G\setminus\{k\}]}$ for all $k\in G$ involving only one direct calculation which we provide below. We have
\beqn
&&\hspace{-0.7cm}F_c^{[G\setminus \{k\}]} (\kan)
=
\tr \left[\left(\tr_{k,k+\gej}  \id _{\ch_{1,2,\dots,\gej}}\otimes \kan _{\ch _{\gej+1,\gej+2,\dots,2\gej}} \left(\bigotimes _{i\in  G} \pplus{i,i+\gej}\right)\right)\right.
\hspace{-0.1cm}\times\non  &&\hspace{-0.7cm}\times\left.\hspace{-0.1cm}\left(\bigotimes _{i\in G \setminus \{k\}} \pplus{i,i+\gej}\right)\hspace{-0.1cm}\right]
= \tr \hspace{-0.15cm}\left[\hspace{-0.1cm}\left(   \id _{\ch_{1,2,\dots,\gej}}\otimes \kan _{\ch _{\gej+1,\gej+2,\dots,2\gej}}  \left(\bigotimes _{i\in G} \pplus{i,i+\gej}\right)\hspace{-0.1cm}\right)\right.\hspace{-0.15cm}\times\non  && \hspace{6.5cm}\times\left. \left(\bigotimes _{i\in G \setminus \{k\}} \pplus{i,i+\gej}\otimes \jed{k,k+\gej}\right)\right],\nonumber
\eeqn
where we have used the property $\tr A \varrho_1=\tr  A^1\otimes\jedynka ^2\varrho_{12}$. Decomposition of $\kan$ into its Kraus components, application of the previously used properties allow us to write
\beqn
F_c^{[G\setminus \{k]\}}(\kan)\hspace{-0.3cm}&=&\hspace{-0.3cm}
\frac{1}{\Pi _{i\in G\setminus\{k\}} d_{i} ^2 d_{k}}\sum_ K\tr \hspace{-0.2cm}\left[\jed{1,2,\dots,\gej}\otimes \akdrugi{\gej} \hspace{-0.1cm}\left(\bigotimes _{i\in G} \swap{i,i+\gej}\right)
\qquad\right.\non &&\hspace{-1.4cm}\left.\times\quad\jed{1,2,\dots,\gej}\otimes \akddrugi{\gej}\left(\bigotimes _{i\in G\setminus \{k\}} \swap{i,i+\gej}\otimes \jed{k,k+\gej}\right)\right].
\nonumber
\eeqn
Finally, inserting identity divided into swaps before the last term under the trace gives
\beqn
F_c^{[G\setminus \{k]\}} (\kan)\hspace{-0.1cm}=\hspace{-0.1cm}
\frac{\sum_ K\tr  \akd{\gej} \otimes \akdrugi{\gej} \hspace{-0.1cm}\left(\jed{G^2,\setminus \{k,k+\gej\}}\otimes \swap{k,k+\gej}\right)}{\Pi _{i\in G\setminus\{k\}} d_{i} ^2 d_{k}}.\nonumber
\eeqn
Consequently, these group fidelities give rise to the terms with only one swap in (\ref{ogolnyrozpis}).
We apply previously described procedure to the remaining terms besides the last one which is easily found to be equal to $\Pi_{i\in G}(d_i+1)^{-1}$.
All above results give us  Eq. (\ref{averagechannel}).

Within the same method group fidelities analogs
can be obtained.


\begin{thebibliography}{}
\bibitem{uwaga} This approach (see \cite{HoHoHo00,KreWer03}) is somewhat more general than
requiring $\mathcal{H}^{(n)}=\mathcal{H}^{\otimes n}$.
\bibitem{simplify} To simplify notation we omit superscript $(n)$ in Appendices.
\end{thebibliography}
\end{document}